\documentclass[12pt]{article}

\usepackage{scicite}

\usepackage{times}

\usepackage{amsfonts}
\usepackage{graphicx}
\usepackage{epsfig}
\usepackage{epsf}
\usepackage{amssymb}
\usepackage{amsmath}
\usepackage{amsthm}
\usepackage{multirow}

\newcommand{\be}{\begin{equation}}
\newcommand{\ee}{\end{equation}}
\newcommand{\bea}{\begin{eqnarray}}
\newcommand{\eea}{\end{eqnarray}}

\newcommand{\la}{\langle}
\newcommand{\ra}{\rangle}
\newcommand{\nl}{\nonumber \\}

\newenvironment{sciabstract}{%
\begin{quote} \bf}
{\end{quote}}

\newcounter{lastnote}

\title{Simulation of Chemical Reaction Dynamics on an NMR Quantum Computer}

\author
{Dawei Lu,$^{1}$ Nanyang Xu,$^{1}$ Ruixue Xu,$^{1}$ Hongwei Chen,$^{1}$\\
Jiangbin Gong,$^{2,3}$ Xinhua Peng,$^{1}$ Jiangfeng Du,$^{1\ast}$\\ \\
 \hspace{-1cm}\normalsize{$^{1}$Hefei National Laboratory for Physical Sciences at
Microscale and Department of Modern Physics,} \\
\hspace{-1cm}\normalsize{University of Science
and Technology of China,}
\normalsize{Hefei, Anhui 230026, People's Republic of
China}\\
\hspace{-1cm}\normalsize{$^{2}$ Department of Physics and Centre for
Computational Science and Engineering,} \\
\normalsize{National
University of Singapore, 117542, Republic of
Singapore} \\
\hspace{-1cm}\normalsize{$^{3}$ NUS Graduate School for Integrative
Sciences and Engineering, 117597,
Republic of Singapore}\\
\hspace{-1cm}\normalsize{$^\ast$To whom correspondence should be addressed; E-mail:  djf@ustc.edu.cn} }

\date{}

\begin{document}


\baselineskip24pt


\maketitle


\begin{sciabstract}
Quantum simulation can beat current classical computers with minimally a few tens of qubits and will likely
become the first practical use of a quantum computer. One promising application of quantum simulation is to attack
challenging quantum chemistry problems.  Here we report an experimental demonstration that a small
nuclear-magnetic-resonance (NMR) quantum computer is already able to
simulate the dynamics of a prototype chemical reaction.
The experimental results agree well with classical simulations.
We conclude that the quantum simulation of chemical reaction dynamics not computable on current classical computers is feasible in the near future.

\end{sciabstract}

\clearpage


\paragraph*{Introduction.}

In addition to offering general-purpose quantum algorithms with substantial speed-ups over classical algorithms \cite{Nielsen}
[e.g., Shor's quantum factorizing algorithm \cite{Shor}], a quantum computer can be used to simulate
specific quantum systems with high efficiency \cite{Buluta}.
This quantum simulation idea was first conceived by Feynman \cite{Feynman}.
Lloyd proved that with quantum computation architecture,
the required resource for quantum simulation scales polynomially
with the size of the simulated system \cite{Lloyd}, as compared with the exponential scaling on  classical computers.
During the past years several quantum simulation algorithms designed for individual problems were
proposed \cite{Zalka,Abrams,Wu,Smirnov,Lidar} and a part of them have been realized
using physical systems such as NMR \cite{Peng,Somaroo,Negrevergne} or trapped-ions \cite{Friedenauer}.
For quantum chemistry problems, Aspuru-Guzik {\it et al.} and Kassal {\it et al.} proposed quantum simulation algorithms
to calculate stationary molecular properties \cite{static} as well as chemical reaction rates \cite{dynamical}, with the quantum simulation
of the former experimentally implemented on both
NMR \cite{static_exp2} and photonic quantum computers \cite{static_exp1}.
In this work we aim at the quantum simulation of the more challenging side of quantum chemistry problems -- chemical reaction dynamics,
presenting an experimental NMR implementation for the first time.

Theoretical calculations of chemical reaction dynamics
play an important role in understanding reaction mechanisms and in guiding
the control of chemical reactions \cite{rabitz,rice-brumer}. On classical computers
the computational cost for propagating the Schr\"{o}dinger equation
increases exponentially with the system size.
Indeed, standard methods in studies of chemical reaction dynamics
so far have dealt with up to 9 degrees of freedom (DOF) \cite{ninedegree}.
Some highly sophisticated approaches,
such as the multi-configurational time-dependent Hartree (MCTDH) method \cite{mctdh},
can treat dozens of DOF but various approximations are necessary.
So generally speaking, classical computers are unable to perform dynamical simulations for large molecules.
For example, for a 10--DOF system
and if only 8 grid points are needed for the coordinate representation of each DOF,
classical computation will have to store and operate $8^{10}$ data points,
already a formidable task for current classical computers.
By contrast, such a system size is manageable by a quantum computer with
only 30 qubits. Furthermore, the whole set of data can be processed in parallel
in quantum simulation.

In this report we demonstrate that the quantum dynamics
of a laser-driven hydrogen transfer model reaction can be captured by
a small NMR quantum simulator.
Given the limited number of qubits, the potential energy curve is modeled by 8 grid points.
The continuous reactant-to-product transformation observed in our quantum simulator is in remarkable agreement
with a classical computation based also upon an 8-dimensional Hilbert space.
To our knowledge, this is the first explicit implementation
of the quantum simulation of a chemical reaction process.  Theoretical methods and general experimental techniques described in this work should motivate
next-generation simulations of chemical reaction dynamics with a larger number of qubits.

\paragraph*{Theory.}

Previously we were able to simulate the ground-state energy of Hydrogen molecule \cite{static_exp2}.
Here, to simulate chemical reaction dynamics, we consider a one-dimensional model of a laser-driven chemical reaction \cite{hsubway}, namely,
the isomerization reaction of nonsymmetric substituted malonaldehydes, depicted in (Fig. 1A).
The system Hamiltonian in the presence of
an external laser field is given by
\be\label{totH}
  H(t)=  T+  V+  E(t) { \quad   \rm  with   \quad    }
                            E(t)=- \mu\varepsilon(t).
\ee
In (Eq. 1),  $E(t)$ is the laser-molecule interaction Hamiltonian,
$ \mu=e  q$ is the dipole moment operator,
$\varepsilon(t)$ represents the driving electric field,
$ {T}={  p}^2/2m$ is the kinetic energy operator, and
\be\label{potential}
 {V}=\frac{\Delta}{2q_0}(  q-q_0)+\frac{V^\ddag-\Delta/2}{q_0^4}(  q-q_0)^2(  q+q_0)^2
\ee
is a double-well potential of the system along the reaction coordinate. In (Eq. 2)
$V^\ddag$ is the barrier height, $\Delta$ gives the asymmetry of the two wells, and $\pm q_0$ give the locations of the potential well minima.
See the figure caption of (Fig. 1B) for more details of this model.

We first employ the split-operator method \cite{Feit,Zhangbook} to obtain the
propagator $ {U}(t+\delta t,t)$ associated with the time interval from $t$ to $t+\delta t$.  We then have
\begin{align}\label{propagator}
 {U}(t+\delta t,t)\approx &\,
 e^{-\frac{i}{\hbar} {V} \delta t/2} e^{-\frac{i}{\hbar} {E} (t+\delta t/2)  \delta t/2}
 e^{-\frac{i}{\hbar} {T} \delta t}   \nl & \times e^{-\frac{i}{\hbar} {E} (t+\delta t/2)           \delta t/2}
 e^{-\frac{i}{\hbar} {V} \delta t/2} .
\end{align}
The unitary operator $e^{-i {T}\delta t/\hbar}$ in (Eq. 3) is diagonal in the momentum representation whereas all the other operators
are unitary and diagonal in the coordinate representation. Such $ {U}(t+\delta t,t)$ can be simulated in a rather simple fashion
if we work with both representations and make transformations between them by quantum Fourier transform (QFT) operations.
To take snapshots of the dynamics we divide
the reaction process into 25 small time steps,
with $\delta t=1.5\,$fs and the total duration $t_f=37.5\,$fs.
The electric field of an ultrashort strong laser pulse is chosen as
\be
  \varepsilon(t)=\left\{
    \begin{array}{cc}
       \varepsilon_0\sin^2(\frac{\pi t}     {2s_1})         ;&\qquad   0\leq t\leq s_1\\
       \varepsilon_0                                        ;&\qquad   s_1<t<s_2\\
       \varepsilon_0\sin^2[\frac{\pi(t_f-t)}{2(t_f-s_2)}]   ;&\qquad   s_2\leq t\leq t_f
    \end{array}
  \right.
\ee
with $s_1=5\,$fs and $s_2=32.5\,$fs.
More details, including an error analysis of the split-operator technique, are given in the supplementary material.
The reactant state at $t=0$ is assumed to be
the ground-state $\left\vert \phi_{0} \right\rangle$ of the bare Hamiltonian $T+V$, which is mainly localized in the left potential well.
The wavefunction of the reacting system at later times is denoted by $\left\vert \psi({t}) \right\rangle$.
The product state of the reaction is taken as the first excited state $\left\vert \phi_{1} \right\rangle$ of $T+V$,
which is mainly localized in the right potential well.

With the system Hamiltonian, the initial reactant state, the product state, and the propagation method outlined above, the next step is
to encode the time-evolving wavefunction $\left\vert \psi({t}) \right\rangle$ and the $T$, $V$, $E(t)$ operators by \emph{n} qubits. To that end
we first obtain the expressions of these operators in representation of a set of $N=2^n$ discretized position basis states.
The evolving state can then be encoded as
\begin{align}\label{wave}
\left\vert \psi(t) \right\rangle=&\sum_{q=0}^{2^n-1}m_q(t)\left\vert q \right\rangle
\nl =&m_0(t)\left\vert 0\cdots00 \right\rangle+...+m_{2^n-1}(t)\left\vert 1\cdots11 \right\rangle,
\end{align}
and as a result the system operators become
\begin{eqnarray}\label{qft}
 {T}&\!\!=&\!\!\sum_{k_1,\cdots ,k_n=z,i}\alpha_{k_1\cdots k_n}\sigma_{k_1}^1\sigma_{k_2}^2 \cdots \sigma_{k_n}^n,\\
 {V}&\!\!=&\!\!\sum_{k_1,\cdots ,k_n=z,i}\beta_{k_1\cdots k_n}\sigma_{k_1}^1\sigma_{k_2}^2 \cdots \sigma_{k_n}^n, \\
 q &\!=&\! \sum_{j=1}^n\gamma_j\sigma_z^j,
\end{eqnarray}
where $\sigma_z^j$ $(j=1,2,\cdots,n)$ is the Pauli matrix and $\sigma_i^j$ is the \emph{N}-dimensional identity matrix.
Because our current quantum computing platform can only offer a limited number of qubits
and the focus of this work is on an
implementation of the necessary gate operations under the above encoding, we have employed
a rather aggressive 8-point discretization using $n=3$ qubits.  The associated diagonal forms of the
$T$, $V$, and $q$ matrices are
given in the supplementary material. In particular,
the end grid points are at $q=\pm 0.8$ \AA\, and the locations of other 6 grid points are shown in (Fig. 1B).
The eigenvalues of the ground and first excited states of the bare Hamiltonian
treated in the 8-dimensional encoding Hilbert space are close to the exact answers. The
associated eigenfunctions are somewhat deformed from exact calculations using, e.g., 64 grid points.
Nonetheless, their unbalanced probability distribution in the two potential wells is maintained.
For example, the probability for the first excited state being found in the right potential well is about $80\%$.




\paragraph*{Experiment.}

In our experiment qubits 1,2, and 3 are realized by the $^{19}$F, $^{13}$C, and $^1$H nuclear spins of Diethyl-fluoromalonate.  The structure of Diethyl-fluoromalonate is shown in (Fig. 2A), where the three nuclei used as qubits are marked by oval. The internal Hamiltonian of this system is given by
\begin{eqnarray}\label{Hamiltonian}
\mathcal{H}_{int}=&&\sum\limits_{j=1}^3 {2\pi \nu _j } I_z^j  + \sum\limits_{j < k,=1}^3 {2\pi} J_{jk} I_z^j I_z^k,
\end{eqnarray}
where $\nu_j$ is the resonance frequency of the \emph{j}th spin and
$\emph{J}_{jk}$ is the scalar coupling strength between spins \emph{j} and
\emph{k}, with $\emph{J}_{12}=-194.4$ Hz, $\emph{J}_{13}=47.6$ Hz, and $\emph{J}_{23}=160.7$ Hz.
The relaxation time $T_1$ and dephasing time $T_2$ for each of the three nuclear spins are tabulated in (Fig. 2B).  The experiment is conducted on a Bruker Avance 400 MHz spectrometer at room temperature.

The experiment consists of three parts: (A) Initial state preparation. In this part we prepare the ground state $\left\vert \phi_{0} \right\rangle$ of the bare Hamiltonian $T+V$ as the reactant state; (B) Dynamical evolution, that is, the explicit implementation of the system evolution such that the continuous chemical reaction dynamics can be simulated; (C) Measurement. In this third part the probabilities of the reactant and product states associated with each of the 25 snapshots of the dynamical evolution are recorded.  For the $j$th snapshot at $t_j\equiv j\delta t$, we measure the overlaps $C(\left\vert \psi(t_j) \right\rangle,\left\vert \phi_{0} \right\rangle)=| \la\phi_0|\psi(t_j)\ra |^2$ and $C(\left\vert \psi(t_j) \right\rangle,\left\vert \phi_{1} \right\rangle)=
|\la\phi_1|\psi(t_j)\ra |^2$, through which the continuous reactant-to-product transformation can be displayed. The main experimental details are as follows. Readers may again refer to the supplementary material for more technical explanations.

(A) $\emph{Initial State Preparation}$. Starting from the thermal equilibrium state, firstly we create the pseudo-pure state (PPS) $\rho_{000}=(1-\epsilon)\mathbb{{I}}/8+\epsilon \left\vert 000 \right\rangle \left\langle000\right\vert$ using the spatial average technique \cite{spatial}, where $\epsilon \approx 10^{-5}$ represents the polarization of the system and ${\mathbb{{I}}}$ is the $8\times 8$ identity matrix.
The initial state $\left\vert \phi_{0} \right\rangle$ was prepared from $\rho_{000}$ by applying one shaped radio-frequency
(RF) pulse characterized by 1000 frequency
segments and determined by the GRadient Ascent Pulse Engineering (GRAPE) algorithm \cite{grape1,grape2,grape3}. The preparation pulse thus obtained
is shown in (Fig. 2C) with the pulse width chosen as 10 ms and a theoretical fidelity 0.995.  Because the central resonance frequencies of the nuclear spins are different, (Fig. 2C) shows the RF field amplitudes vs time in three panels.
To confirm the successful preparation of the state $\vert\phi_0\rangle$,   we carry out
 a full state tomography and examine the fidelity between the target density matrix $\rho_0=|\phi_0\rangle\langle\phi_0|$ and the experimental one $\rho_{exp}(0)$.  Using the
 fidelity definition $F(\rho_{1}, \rho_{2})\equiv \texttt{Tr}(\rho_1{\rho_2})/\sqrt{(\texttt{Tr}(\rho_1^2)\texttt{Tr}(\rho_2^2)}$,
 we obtain $F[\rho_0, \rho_{exp}(0)]=0.950$. Indeed, their real parts shown in (Fig. 4A) are seen to be in agreement.

(B) $\emph{Dynamical Evolution}$. The reaction process was divided into $M=25$ discrete time intervals of the same duration $\delta t$.
Associated with the $m$th time interval, the unitary evolution operator is given by
\begin{equation}
U_m\approx V_{\delta t/2}{E}_{\delta t/2}(t_m)U_{QFT}T_{\delta t}U_{QFT}^{\dagger}{E}_{\delta t/2}(t_{m}){V}_{\delta t/2},
\end{equation}
where $U_{QFT}$ represents a QFT operation, and other operators are defined by ${V}_{\delta t/2}\equiv e^{-\frac{i}{\hbar}{V}\frac{\delta t}{2}}$, ${T}_{\delta t} \equiv e^{-\frac{i}{\hbar}{T}\delta t}$, and ${E}_{\delta t/2}(t_m)\equiv e^{\frac{i}{\hbar}\varepsilon(t_{m-1}+\delta t/2) e q\frac{\delta t}{2}}$, with $V$, $T$, and $q$ all in their diagonal representations.  Such a loop of operations is $m$-dependent because the simulated system is subject to a time-dependent laser field.  The numerical values of the diagonal operators $T_{\delta t}$, $V_{\delta t/2}$ and $E_{\delta t/2}$ are
elaborated in the supplementary material.  A circuit to realize $U_{QFT}$ and a computational network to realize the $U_m$ operator
are shown in (Fig. 3).

Each individual operation in the $U_m$ loop  can be implemented by a particular RF pulse sequence applied to our system. However, in the experiment such a direct decomposition of $U_m$ requires a very long gate operation time and highly complicated RF pulse sequences. This bottom-up approach hence accumulates considerable experimental errors and
also invites serious decoherence effects. To circumvent this technical problem we find a better experimental approach, which further
exploits the GRAPE technique to synthesize $U_m$ or their products with one single engineered RF pulse only. That is, the quantum evolution operator $U(t_j, 0)$, which is simulated by $\prod_{m=1}^{j}U_m$, is implemented by one GRAPE coherent control pulse altogether, with a preset fidelity and a typical pulse length ranging from 10 ms to 15 ms.  For the 25 snapshots of the dynamics, totally 25 GRAPE pulses are worked out, with their fidelities always set to be larger than 0.99.  As a result, the technical complexity of the experiment decreases dramatically but the fidelity is maintained at a high level.  The task of finding a GRAPE pulse itself may be fulfilled via feedback learning control \cite{rabitz} that can exploit the quantum evolution of our NMR system itself.  However, this quantum procedure is not essential or necessary in our experiment because here the GRAPE pulses on a 3-qubit system can be found rather easily.

(C) $\emph{Measurement}$. To take the snapshots of the reaction process at $t_j=j\delta t$ we need to measure the overlaps of C($\left\vert \psi(t_j) \right\rangle$,$\left\vert \phi_{0} \right\rangle$) and C($\left\vert \psi({t_j}) \right\rangle$,$\left\vert \phi_{1} \right\rangle$). A full state tomography at $t_j$ will do, but this will produce much more information than needed. Indeed, assisted by a simple diagonalization technique, sole population measurements already suffice to observe the reactant-to-product transformation.

Specifically, in order to obtain $C(\left\vert \psi({t_j}) \right\rangle,\left\vert \phi_{0} \right\rangle)=\texttt{Tr}[\rho(t_j) \rho_0]$ with $\rho(t_j)=\left\vert \psi({t_j}) \right\rangle\left\langle \psi(t_j) \right\vert$, we first find a transformation matrix \emph{R} to diagonalize $\rho_0$, that is, $\rho_0'=R \rho_0R^{\dagger}$, where $\rho_0'$ is a diagonal density matrix. Letting $\rho'(t_j) = R \rho(t_j) R^{\dagger}$ and using the identity $\texttt{Tr}[\rho(t_j) \rho_0]=\texttt{Tr}[\rho'(t_j) \rho_0']$, it becomes clear
that only the diagonal elements or the populations of $\rho'(t_j)$ are required to measure $\texttt{Tr}[\rho'(t_j) \rho_0']$, namely, the overlap $C(\left\vert \psi(t_j) \right\rangle,\left\vert \phi_{0} \right\rangle)$.  To obtain $\rho'(t_j)$ from $\rho(t_j)$, we simply add the extra $R$ operation to the quantum gate network. The actual implementation of the $R$ operation can be again mingled with all other gate operations using one GRAPE pulse. A similar procedure is used to measure C($\left\vert \psi(t_j) \right\rangle$,$\left\vert \phi_{1} \right\rangle$).

The populations of $\rho{'}(t_j)$ can be measured by applying $[\pi/2]_y$ pulses to the three qubits and then read the ensuing free induction decay signal. In our sample of natural abundance, only $\sim 1\%$ of all the molecules contain a ${}^{13}$C nuclear spin.  The signals from the ${}^{1}$H and ${}^{19}$F nuclear spins are hence dominated by those molecules with the ${}^{12}$C isotope. To overcome this problem we apply SWAP gates to transmit the information of the ${}^{1}$H and ${}^{19}$F channels to the ${}^{13}$C channel and then measure the ${}^{13}$C qubit.  

 To assess the difference between theory and experiment,  we carry out one full state tomography for the final state density matrix  at $t=t_f$.
 Because the GRAPE pulse is made to reach a fidelity larger than 0.995, the experimental density matrix $\rho_{exp}(t_f)$
 is indeed very close to the  theoretical density matrix $\rho_{theory}(t_f)$ obtained in an 8-dimensional Hilbert space, with a fidelity $F[\rho_{theory}(t_f),\rho_{exp}(t_f)]=0.957$. The experimental density matrix elements of the final state shown in (Fig. 4B) match the theoretical results to a high degree. With confidence in the experimental results on the full density matrix level, we can now examine the simulated reaction dynamics, reporting only the probabilities of the reactant and product states.
 (Fig. 4C) shows the time-dependence of the probabilities of both the reactant and product states obtained from our quantum simulator.
 It is seen that the product-to-reactant ratio
  increases continuously with time, with the probability of the product state reaching 77\% at the end of the simulated reaction. At all times,
  the experimental observations of the reaction process are in impressive agreement with the smooth curves calculated theoretically on a classical computer. Further, the experimental results are also in qualitative agreement
   with the exact classical calculation using 64 grid points (see Fig. 1B).
    A prototype laser-driven reaction is thus successfully simulated by our 3-qubit system.
    We emphasize that due to the use of GRAPE pulses in synthesizing the gate operations, our simulation experiment lasts about 30 ms only, which is much shorter than the spin decoherence time of our system.
  The slight difference between theory and experiment can be attributed to imperfect GRAPE pulses, as well as inhomogeneity in RF pulses and in the static magnetic field.

\paragraph*{Conclusion.}

 Quantum simulation with only tens of qubits can already exceed the capacity of a classical computer.  Before realizing general-purpose quantum algorithms that typically require thousands of qubits, a quantum simulator attacking problems not solvable on current classical computers will be one conceivable milestone in the very near future. The realization of quantum simulations will tremendously change the way we explore quantum chemistry in both stationary and dynamical problems \cite{static,dynamical}.   
Our work reported here establishes the first experimental study of the quantum simulation of a prototype laser-driven chemical reaction. The feasibility of simulating chemical reaction processes on a rather small quantum computer is hence demonstrated.  Our proof-of-principle experiment also realizes a promising map from laser-driven chemical reactions to the dynamics of interacting spin systems under shaped RF fields. This map itself is of significance because it bridges up two research subjects whose characteristic time scales differ by many orders of magnitude.



\clearpage

{\bf References and Notes}

\clearpage

\noindent {\bf Fig. 1.}
Prototype chemical reaction and potential energy curve.
(A) Isomerization reaction of nonsymmetric substituted malonaldehydes.
(B) Upper panel: Potential energy curve, together with the eigenfunctions of
    the ground (red) and the first excited (blue) states.
    The main system parameters [(Eq. 2)] are taken from Ref.\ \cite{hsubway}, with
    $V^\ddag=0.00625\ E_{\rm h}$, $\Delta=0.000257\ E_{\rm h}$, and $q_0=1\ a_0$.
    As a modification, the potential values for $q$ approaching the left and right ends
are increased sharply to ensure rapid decay
    of the wavefunction amplitudes. In particular, this procedure increases
    the $V$ value at $q=\pm 0.8$ \AA\ by a factor of 30.
    The six discrete squares shown on the potential curve and the two end points at $q=\pm 0.8$ \AA
    \ constitute the 8 grid points for our 3-qubit encoding.
    Lower panel: Numerically exact time-dependence of populations of the ground state (reactant state, denoted P$_0$) and the first excited state
    (product state, denoted P$_1$).

\clearpage

\noindent {\bf Fig. 2.}
(A) Molecular structure of Diethyl-fluoromalonate.
The $^1$H, $^{13}$C and $^{19}$F nuclear spins marked by oval are used as three qubits.
(B) System parameters and important time scales of Diethyl-fluoromalonate.  Diagonal elements are the Larmor frequencies (Hz) and
off-diagonal elements are scalar coupling strength (Hz) between two nuclear spins.
Relaxation and dephasing time scales (second) $T_1$ and $T_2$ for each nuclear spin are listed on the right.
(C) The GRAPE pulse that realizes the initial state $\left\vert \phi_{0} \right\rangle$ from the PPS $\left\vert 000 \right\rangle$, with a pulse width 10 ms and a fidelity over 0.995. The (blue)
solid line represents the pulse power in $x$-direction, and the (red)
dotted line represents the pulse power in $y$-direction. The three panels from top to bottom represent the RF features at three central frequencies
associated with
the $^{19}$F,  $^{13}$C and $^1$H spins, respectively.

\clearpage

\noindent {\bf Fig. 3.}
Upper panel: The network of quantum operations to simulate the chemical reaction dynamics, with the reactant state $\left\vert \phi_{0} \right\rangle$.
The whole process is divided into 25 loops. The operators $T_{\delta t}$, $V_{\delta t}$ and $E_{\delta t/2}$ are assumed to be in their
diagonal representations.  Lower panel:
H is the Hadamard gate and S, T are phase gates as specified on the right. Vertical lines ending with a solid dot represent controlled phase gates and the
vertical line between two crosses represents a SWAP gate.

\clearpage

\noindent {\bf Fig. 4.}
Experimental tomography results and the reaction dynamics obtained both theoretically and experimentally.
(A)-(B) Real part of the density matrix of the initial and final states of the simulated reaction.
Upper panels show the theoretical results based on an 8-dimensional Hilbert space, and lower panels show the experimental results.
(C) The measured probabilities of the reactant and product states to give 25 snapshots of the reaction dynamics.   The (red) plus symbols represent measured results of C($\left\vert \psi(t_j)\right\rangle$,$\left\vert \phi_{0} \right\rangle$) and the (blue) circles represent measured results of C($\left\vert \psi(t_j) \right\rangle$,$\left\vert \phi_{1} \right\rangle$), both in agreement with the theoretical smooth curves.  Results here also agree qualitatively
with the numerically exact dynamics shown in (Fig. 1B).

\clearpage
 \begin{center} Fig. 1: \end{center}
\begin{figure}[h]\centering
\includegraphics[width=11cm]{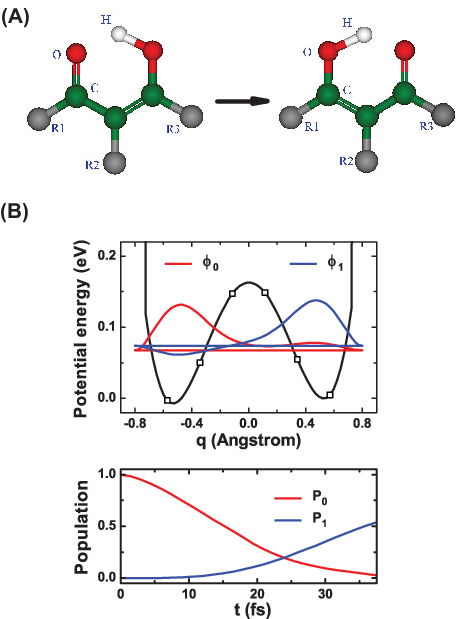}
\label{fig1}
\end{figure}

\clearpage
\begin{center} Fig. 2: \end{center}
\begin{figure}[h]\centering
\includegraphics[width=13cm]{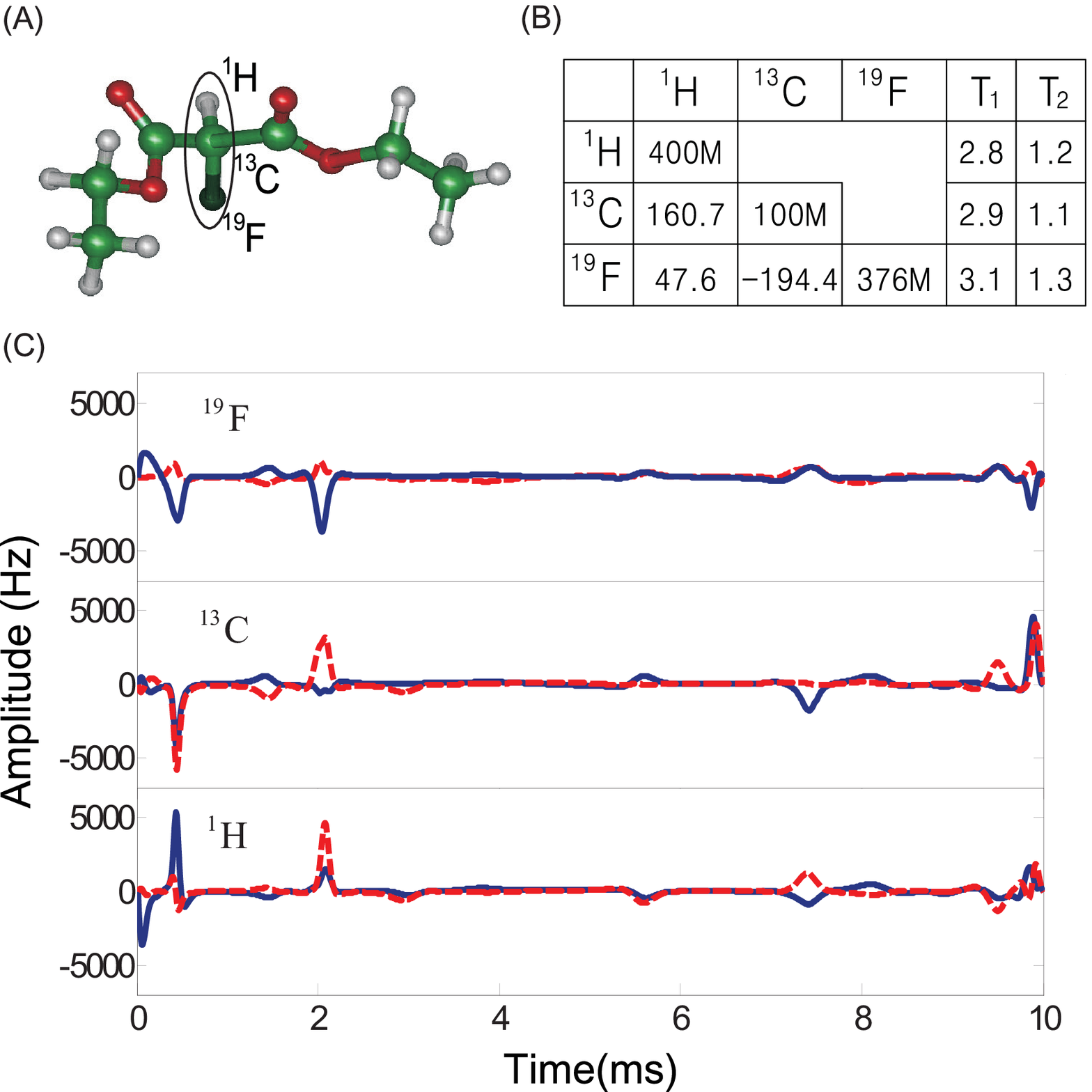}
\label{structure}
\end{figure}

\clearpage
\begin{center} Fig. 3: \end{center}
\begin{figure}[h]\centering
\includegraphics[width=12cm]{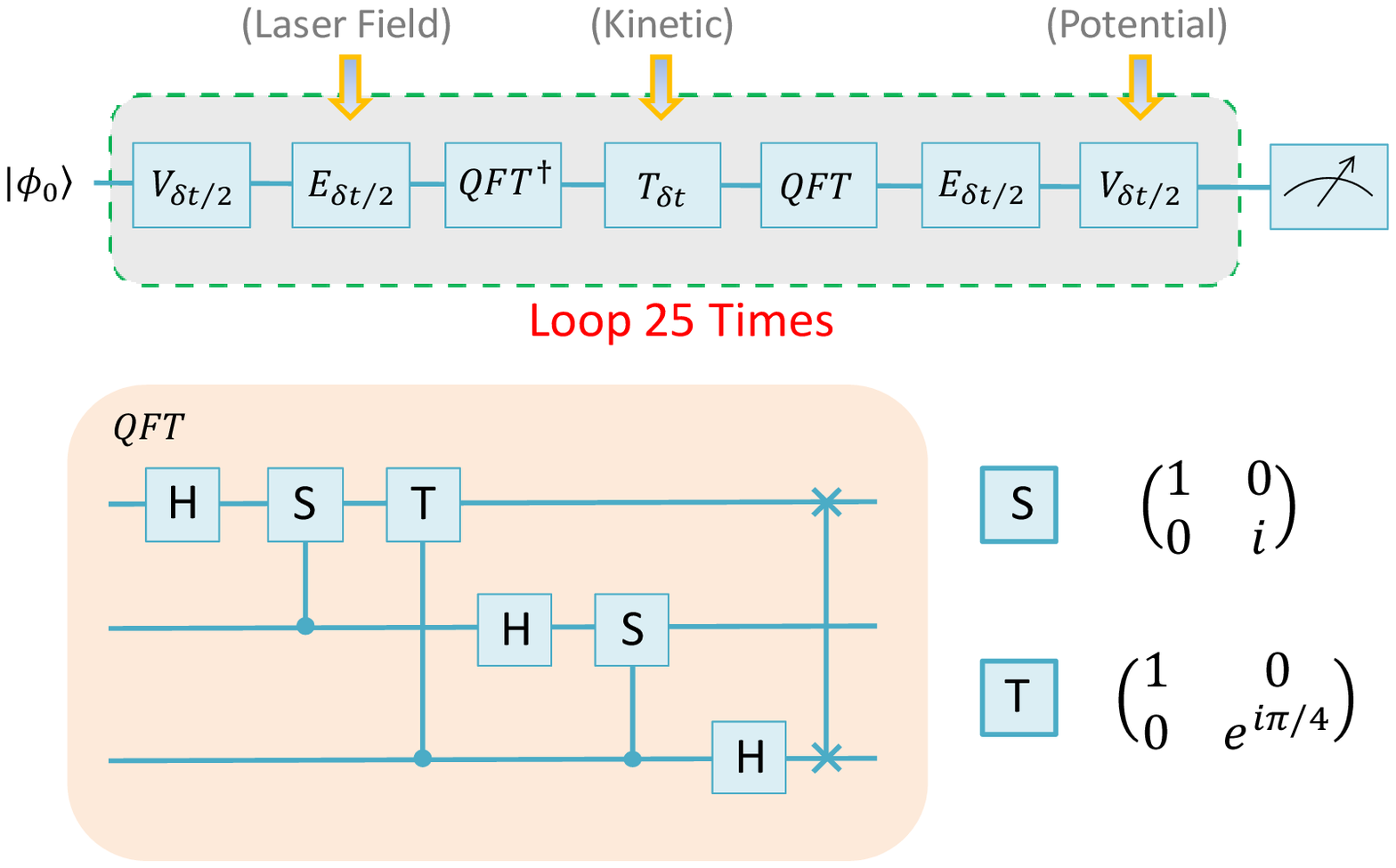}
\label{circuit}
\end{figure}

\clearpage
\begin{center}Fig. 4:\end{center}
\begin{figure}[h]\centering
\includegraphics[width=11cm]{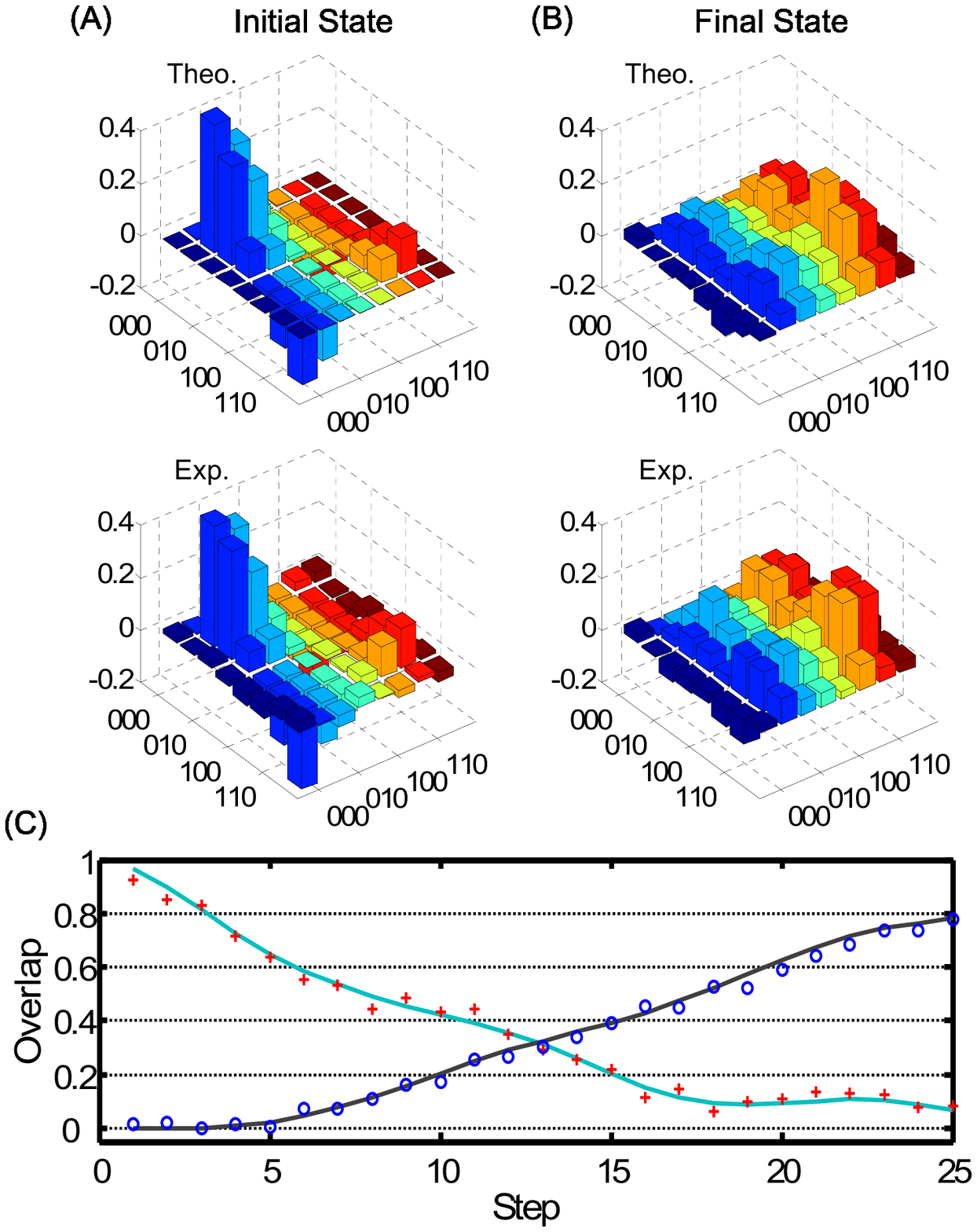}
\label{tomo}
\end{figure}

\clearpage

\section*{Supporting Online Material}

\subsection*{BACKGROUND ON QUANTUM DYNAMICS SIMULATION}

Let us start with the Schr\"{o}dinger equation:
\be
 i\hbar \dot\psi(t)=  H(t)\psi(t).
\ee
Its formal solution can be written as
\be
   \psi(t)=   U(t,t_0)\psi(t_0),
\ee
where the quantum propagator $U(t,t_0)$ is a unitary operator and is given by
\be
     U(t,t_0)={\cal T}\exp\left[-\frac{i}{\hbar}
  \int_{t_0}^t\!\!d\tau\,   H(\tau)\right],
\ee
with ${\cal T}$ being the time ordering operator.  There are a number of established numerical methods for propagating the Schr\"{o}dinger equation,
such as Feynman's path integral formalism \cite{Feynman2},
the split-operator method \cite{Feit2} and the Chebychev polynomial method \cite{Kosloff,Zhangbook2}, etc.  For our purpose here we adopt the split-operator method.

The propagator $U(t,t_0)$  satisfies
\be
     U(t,t_0)=   U(t,t_{N-1})   U(t_{N-1},t_{N-2})\cdots\cdots   U(t_1,t_0)\,,
\ee
where, for example, the intermediate time points can be equally spaced with $t_m=m\delta t+t_0$.
For one of such a small time interval, e.g., from $t_{m-1}$ to $t_{m}=t_{m-1}+\delta t$, we have
\begin{eqnarray}
     U(t_{m},t_{m-1})& = &{\cal T} \exp\left[-\frac{i}{\hbar}
  \int_{t_{m-1}}^{t_m}\!\!d\tau\,   H(\tau)\right] \nonumber \\
  &\approx & \exp\left[-\frac{i}{\hbar}
  \int_{t_{m-1}}^{t_m}\!\!d\tau\,   H(\tau)\right],
\end{eqnarray}
where terms of the order $(\delta t)^3$ or higher are neglected.  For sufficiently small $\delta t$,
the integral in the above equation can be further carried out by a midpoint rule, leading to
\begin{eqnarray}
     U(t_{m},t_{m-1})\approx \exp\left[-\frac{i}{\hbar} H(t_{m-1}+\delta t/2) \delta t\right].
     \label{someq}
\end{eqnarray}
This integration step has an error of the order of $(\delta t)^2$, which is still acceptable if the total evolution time
is not large.
Next we separate the total Hamiltonian into two parts:
\be
     H(t)=   H_0(t)+   H'(t).
\ee
For example, $H_0(t)$ is the kinetic energy part of the total Hamiltonian and $H'(t)$ represents the potential energy part.  In general
$H_0(t)$ and $H'(t)$ do not commute with each other.
The split-operator scheme \cite{Feit2}  applied to (Eq. \ref{someq}) then leads to
\be
  {U}(t_m,t_{m-1})\approx
 e^{-\frac{i}{\hbar}   H' (t_{m-1}+\delta t/2)    \delta t/2}
 e^{-\frac{i}{\hbar}   H_0  (t_{m-1}+\delta t/2)  \delta t  }
 e^{-\frac{i}{\hbar}   H' (t_{m-1}+\delta t/2)             \delta t/2}.
 \label{sop}
\ee
The small error of this operator splitting step arises from the nonzero commutator between
$ H_0(t)$ and $H'(t)$, which is at least of the order of $(\delta t)^3$.  The advantage of
the split-operator method is that each step represents a unitary evolution and each exponential in (Eq. \ref{sop}) can take a diagonal form
in either the position or the momentum representation.
  The error of this operator splitting step
is in general smaller than that induced by the aforementioned midpoint rule integration in
(Eq. \ref{someq}). Because in our work the total duration of the
simulated chemical reaction is short, the above low-order approach already has a great performance.
If long-time simulations with preferably larger time steps are needed in a quantum simulation,
then one may use even higher-order split-operator techniques for explicitly time-dependent Hamiltonians \cite{bandrauk,zhu}.

\subsection*{EXPERIMENTAL IMPLEMENTATION}

The experiment consists of three steps: (a) Initial state preparation, which is to prepare the ground state $\left\vert \phi_{0} \right\rangle$ of the bare Hamiltonian $T+ {V}$ (representing the reactant state); (b) 25 discrete steps of dynamical evolution to simulate the actual continuous chemical reaction dynamics; (c) Measurement of the overlaps $C(\left\vert \psi(t_j) \right\rangle,\left\vert \phi_{0} \right\rangle)=| \la\phi_0|\psi(t_j)\ra |^2$ and $C(\left\vert \psi(t_j) \right\rangle,\left\vert \phi_{1} \right\rangle)=| \la\phi_1|\psi(t_j)\ra |^2$ at $t_j=j\delta t$, which is to show the transformation between the reactant and product states.

\subsubsection*{\textbf{A. Initial State Preparation}}

To prepare the ground state $\left\vert \phi_{0} \right\rangle$, firstly we need to create a pseudo-pure state (PPS) from the thermal equilibrium state - a mixed state which is not yet ready for quantum computation purposes. The thermal equilibrium state of our sample can be written as $\rho_{ther}=\sum\limits_{i=1}^3 \gamma_i I_z^i$,
where $\gamma_i$ is the gyromagnetic ratio of the nuclear spins. Typically, $\gamma_\texttt{C}=1$, $\gamma_\texttt{H}=4$ and $\gamma_\texttt{F}=3.7$, with a constant factor ignored. We then use the spatial average technique \cite{spatial-2} to prepare the PPS
\begin{equation}\label{ppsform}
\rho_{000}=\frac{1-\epsilon}{8}\mathbb{{I}}+\epsilon \left\vert 000 \right\rangle \left\langle000\right\vert,
\end{equation}
where $\epsilon \approx 10^{-5}$ represents the polarization of the system and ${\mathbb{{I}}}$ is the $8\times8$ unity matrix. The unity matrix
has no influence on our experimental measurements and hence can be dropped.
 The pulse sequence to prepare the PPS from the thermal equilibrium state is shown in (Fig. S1A). In particular,
 the gradient pulses (represented by G$_\text{Z}$) destroys the coherence induced by the rotating pulses and free evolutions. After obtaining the PPS,
 we apply one shaped pulse calculated by the GRadient Ascent Pulse Engineering (GRAPE) algorithm \cite{grape1-2,grape2-2,grape3-2} to obtain the initial state $\left\vert \phi_{0} \right\rangle$, with the pulse width 10 ms and a fidelity 0.995. In order to assess the accuracy of the experimental preparation of the initial state, a full state tomography \cite{tomography} is implemented. The fidelity \cite{fidelity} between the target density matrix $\rho_{target}$ and the experimental density matrix $\rho_{exp}$ is found to be
\begin{eqnarray}\label{pps}
F(\rho_{target}, \rho_{exp})&\equiv& \texttt{Tr}(\rho_{target}\rho_{exp})/\sqrt{(\texttt{Tr}(\rho_{target}^2)\texttt{Tr}(\rho_{exp}^2)} \nonumber \\
&\approx & 0.95.
\end{eqnarray}
A detailed comparison between $\rho_{target}$ and $\rho_{exp}$ is displayed in (Fig. S1B).

\subsubsection*{\textbf{B. Dynamical Evolution}}

To observe the continuous reactant-to-product transformation, we divide the whole time evolution into 25 discrete steps. For convenience
all variables here are expressed in terms of atomic units.  For example, in atomic units $e=1$, $\hbar=1$, and $\delta t  = 62.02$.
  To exploit the split-operator scheme in (Eq. \ref{sop}), we let $H_0=T$, and $H'(t)=V-eq\epsilon(t)$.  The kinetic energy operator $T$ is diagonal in the momentum representation, whereas the $V$ operator and the dipole-field interaction $-eq\epsilon(t)$ operator are both diagonal in the position representation.  We then obtain from (Eq. \ref{sop})
\begin{eqnarray}
 U(t_m,t_{m-1})\approx  V_{\frac{\delta t}{2}}E_{\frac{\delta t}{2}}U_{QFT}T_{\delta t}U_{QFT}^{\dagger}E_{\frac{\delta t}{2}}V_{\frac{\delta t}{2}},
 \end{eqnarray}
where the operators
\begin{eqnarray}\label{potential}
V_{\frac{\delta t}{2}} &\equiv &e^{-i {V}\frac{\delta t}{2}},\\
T_{\delta t} &\equiv &e^{-i  {T}\delta t},\\
E_{\frac{\delta t}{2}} &\equiv&e^{i {q}\varepsilon(t_{m-1}+\delta t/2)\frac{\delta t}{2}}
\end{eqnarray}
are assumed to be in their diagonal representations.

To map $U(t_m,t_{m-1})$ to our 3-qubit NMR quantum computer, we discretize the potential energy curve using 8 grid points. Upon this discretization,
operators $V_{\frac{\delta t}{2}}$, $T_{\delta t}$, and $q$ become $8\times 8$ diagonal matrices. Numerically,  their diagonal elements  (denoted
$V_{diag}$, $T_{diag}$, and $q_{diag}$, respectively) are found to be
\begin{eqnarray}\label{operator}
  {V}_{diag} =&&(293.78,-0.10,1.85,5.41,\nonumber\\
             &&  5.46,2.02,0.18,305.44)\times 10^{-3};\nonumber\\
  {T}_{diag} =&&(0,0.91,3.63,8.16,\nonumber\\
&&  14.51,8.16,3.63,-0.91)\times 10^{-3};\nonumber\\
  {q}_{diag} =&&(-1.51,-1.08,-0.65,-0.22,\nonumber\\
&&  0.22,0.65,1.08,1.51).
\end{eqnarray}
To evaluate the $E_{\frac{\delta t}{2}}$ operator, we also need to discretize the time-dependence of the electric field associated with the ultrashort laser pulse.  For 25 snapshots of the reaction dynamics, we discretize the trapezoid-type electric field by 25 points, i.e., \begin{equation}\label{varepsilon}
\varepsilon(t)=[0.05,0.42,0.85,1,...1,0.85,0.42,0.05] \times 10^{-3}.
\end{equation}

The quantum gate network for the QFT operation that transforms the momentum representation to the coordinate representation is already shown in (Fig. 3). It consists of three Hardmard gates (H), three controlled-phase gates (S and T) and one SWAP gate (vertical line linking crosses).
The Hardmard gate H is represented by the Hardmard matrix
\begin{eqnarray}\label{Hardmard}
H = \frac{1}{\sqrt{2}}\left(
  \begin{array}{cc}
    1 & 1 \\
    1 & -1 \\
  \end{array}
\right),
\end{eqnarray}
which maps the basis state $\left\vert 0 \right\rangle$ to $\frac{1}{\sqrt{2}}(\left\vert 0 \right\rangle+\left\vert 1 \right\rangle)$ and $\left\vert 1 \right\rangle$ to $\frac{1}{\sqrt{2}}(\left\vert 0 \right\rangle-\left\vert 1 \right\rangle)$. The phase gates S and T are given by
\begin{eqnarray}\label{phasegate}
\text{S} = \left(
  \begin{array}{cc}
    1 & 0 \\
    0 & i \\
  \end{array}
\right)
\end{eqnarray}
and
\begin{eqnarray}\label{phasegate}
\text{T} = \left(
  \begin{array}{cc}
    1 & 0 \\
    0 & e^{i\pi/4} \\
  \end{array}
\right).
\end{eqnarray}
The matrix form of the SWAP gate is
\begin{eqnarray}\label{SWAP}
\text{SWAP} = \left(
         \begin{array}{cccc}
           1 & 0 & 0 & 0 \\
           0 & 0 & 1 & 0 \\
           0 & 1 & 0 & 0 \\
           0 & 0 & 0 & 1 \\
         \end{array}
       \right),
\end{eqnarray}
which exchanges the state of the $^{19}$F qubit and with that of the $^1$H qubit.

{\it GRAPE Pulses.} Since there are hundreds of logical gates in the required network of quantum operations, a direct implementation of the gate operation network will need a large number of single-qubit rotations as well as many free evolutions during the single-qubit operations. This bottom-up approach will
then accumulate the errors in every single-qubit operation. Considerable decoherence effects will also emerge during the long process. For example,
we have attempted to directly decompose the network into a sequence of RF pulses, finding that the required free evolution time for the 25 loops of evolution is more than 1 s, which is comparable to the $T_2$ time of our system.
To overcome these problems and to reach a high-fidelity quantum coherent control over the three interacting qubits, the unitary operators used in our experiment are realized by shaped quantum control pulses found by the GRadient Ascent Pulse Engineering (GRAPE) technique \cite{grape1-2,grape2-2,grape3-2}.
To maximize the fidelity of the experimental propagator as compared with the ideal gate operations, we use a mean gate fidelity by averaging over a weighted distribution of RF field strengths to minimize the inhomogeneity effect of the RF pulses applied to the sample.

For a known or desired unitary operator $U_{target}$,
the goal of the GRAPE algorithm is to find a shaped pulse within a given duration $t_{total}$ to maximize the fidelity
\begin{equation}\label{grape_fid}
F =|\texttt{Tr}(U_{target}^{\dagger}U_{cal})/2^n|^2,
\end{equation}
where $U_{cal}$ is the unitary operator actually realized by the shaped pulse and $2^n$ is the dimension of the Hilbert space. We discretize the evolution time $t_{total}$ into $N$ segments of equal duration $\Delta t = t_{total}/N$, such that $U_{cal}=U_N\cdots U_2U_1$, with the evolution operator associated with the \emph{j}th time interval given by
\begin{equation}
U_j=\exp\{-i\Delta t(\mathcal{H}_{int}+\sum_{k=1}^m u_k(j)\mathcal{H}_{k})\}.
\end{equation}
Here $\mathcal{H}_{int}$ is the three-qubit self-Hamiltonian in the absence of any control field, $\mathcal{H}_{k}$ represents the interaction Hamiltonians due to the applied RF field, and $u_k(j)$ is the control vectors associated with $\mathcal{H}_{k}$. Specifically, in our experiment $u_k(j)$ are the time-dependent amplitudes of the RF field along the \emph{x} and \emph{y} directions, for the F-channel, the C-channel and the H-channel.  With an initial guess for the pulse shape, we use the GRAPE algorithm to optimize $u_k(j)$ iteratively until $U_{cal}$ becomes very close to $U_{target}$.  More details can be found from Ref. \cite{grape1-2}. The GRAPE technique dramatically decreases the duration and complexity of our experiment and at the same time increases the quantum control fidelity.  In our proof-of-principle demonstration of the feasibility of the quantum simulation of a chemical reaction, the task of searching for the GRAPE pulses is carried out on a classical computer in a rather straightforward manner. It is important to note that this technique can be scaled up for many-qubit systems, because the quantum evolution of the system itself can be exploited in finding the high-fidelity coherent control pulses.

As an example,  (Fig. S2) shows the details of one 15 ms GRAPE pulse to realize the quantum evolution from $t=0$ to $t_7=7\delta t$ (also combining the operations for initial state preparation and the extra operation $R$ that is useful for the measurement stage). The shown GRAPE pulse is found by optimizing the frequency spectrum divided into 750 segments. The shown GRAPE pulse has a fidelity over 0.99.

\subsubsection*{\textbf{C. Measurement}}

To simulate the process of a chemical reaction, it is necessary to measure the simulated reactant-to-product transformation at different times. To that end we measure the overlaps of $C(\left\vert \psi({t_m}) \right\rangle,\left\vert \phi_{0} \right\rangle)$ and $C(\left\vert \psi(t_{m}) \right\rangle,\left\vert \phi_{1} \right\rangle)$ at $t_m=m\delta t$.
Here we first provide more explanations about how a diagonalization method can reduce the measurement of the overlaps to population measurements.

Without loss of generality, we consider the measurement $C(\left\vert \psi_{7} \right\rangle,\left\vert \phi_{0} \right\rangle)$.
\begin{equation}\label{overlap7}
C(\left\vert \psi(t_{7}) \right\rangle,\left\vert \phi_{0} \right\rangle)=| \la\phi_0|\psi(t_7)\ra |^2=\texttt{Tr}[\rho(t_7) \rho_0],
\end{equation}
where $\rho(t_7)=\left\vert \psi(t_{7} \right\rangle\left\langle \psi(t_{7}) \right\vert$ and $\rho_0=\left\vert \phi_{0} \right\rangle\left\langle \phi_{0} \right\vert$. Let \emph{R} be a transformation matrix which diagonalizes $\rho_0$ to a diagonal density matrix $\rho_0'=R \rho_0R^{\dagger}$. Then
\begin{eqnarray}\label{diag}
\texttt{Tr}[\rho(t_7) \rho_0]=\texttt{Tr}[R\rho(t_7) R^{\dagger}R \rho_0R^{\dagger}]=\texttt{Tr}[\rho'(t_7) \rho_0'],
\end{eqnarray}
where $\rho'(t_7)=R \rho(t_7)R^{\dagger}$.
Clearly then, only the diagonal terms (populations) of $\rho'(t_7)$ are relevant when calculating $\texttt{Tr}[\rho'(t_7) \rho_0']$,
namely, the overlap $C(\left\vert \psi(t_{7} \right\rangle,\left\vert \phi_{0} \right\rangle)$. Hence only population measurement of the  density matrix $\rho'(t_7)$ is needed to obtain the overlap between $|\psi(t_7)\rangle$ and the initial state.
The GRAPE pulse that combines the operations for initial state preparation, for the quantum evolution, as well as for the extra operation \emph{R} is shown in (Fig. S2).

 The three population-readout spectra after applying the GRAPE pulse
 are shown in (Fig. S3B-S3D), together with the $^{13}$C spectrum for the PPS $|000\rangle$.
 The populations to be measured are converted to the observable coherent terms by applying a $[\pi/2]_y$ pulse to each of the three qubits.   For the $^{13}$C spectrum shown in (Fig. S3C), four peaks from left to right are seen, with their respective integration results representing $P(5)-P(7)$, $P(6)-P(8)$, $P(1)-P(3)$, and $P(2)-P(4)$, where $P(i)$ is the  \emph{i}th diagonal element of $\rho'(t_7)$.  Experimentally the four integrals associated with the four peaks in (Fig. S3C) are found to be $-0.098$, $-0.482$, $-0.089$ and $-0.071$, which are close to the theoretical values $-0.047$, $-0.501$, $-0.114$ and $-0.041$.
 Further using other readouts from the $^{19}$F (see Fig. S3B) and $^1$H (see Fig. S3D) spectra as well as the normalization condition $\sum_{i=1}^{8} {P}({i})=1$, we obtain all the 8 populations and hence the overlap $C(\left\vert \psi(t_{7} \right\rangle,\left\vert \phi_{0} \right\rangle)$.  The theoretical and experimental results for this overlap are 0.535 and 0.529, which are in agreement.
 A similar procedure is used to obtain $C(\left\vert \psi(t_{m} \right\rangle,\left\vert \phi_{1} \right\rangle)$.

 The spectra of the ${}^{1}$H and ${}^{19}$F channel are obtained by first transmitting the signals of the $^{19}$F and $^1$H qubits to the $^{13}$C qubit using SWAP gates. With this procedure all the spectra shown in (Fig. S3) are exhibited on the $^{13}$C channel.  Indeed, because
 in our sample of natural abundance, only $\approx 1\%$ of all the molecules contain a ${}^{13}$C nuclear spin,
 the signals from the ${}^{1}$H and ${}^{19}$F nuclear spins  without applying SWAP gates would be
  dominated by those molecules with the ${}^{12}$C isotope.

\clearpage

\noindent {\bf Fig. S1.}
Pulse sequence for the preparation of the PPS and a comparison between experimental and theoretical density matrix elements for the reactant state $\left\vert \phi_{0} \right\rangle$.
(A) Pulse sequence that implements the PPS preparation, with $\theta=0.64\pi$ and $\texttt{X}(\overline{\texttt{X}}, \texttt{Y}, \overline{\texttt{Y}})$ representing rotations around the $x$($-x$, $y$, $-y$) direction. G$_\text{Z}$ represents a gradient pulse to destroy the coherence induced by
  the rotating pulses and free evolutions. $\frac{1}{4\text{J}_{\text{HF}}}$ and $\frac{1}{4\text{J}_{\text{CF}}}$ represent the free evolution of the system under $\mathcal{H}_{int}$ for 5.252 ms and 1.286 ms, respectively.
(B) Comparison between measured density matrix elements of the initial state $\left\vert \phi_{0} \right\rangle$ and the theoretial target density matrix elements based on the 8-point encoding. Both the real part and the imaginary part of the density matrix elements are shown.



\clearpage

\noindent {\bf Fig. S2.}
GRAPE pulse to simulate the quantum evolution of the reacting system from $t=0$ to $t_7=7\delta t$.
The top, middle and bottom panels depict the time-dependence of the RF pulses applied to the F-channel, C-channel and H-channel, respectively.
The (blue) solid line represents the pulse power applied in the ${x}$-direction, and the (red) dotted line represents the pulse power applied in the ${y}$-direction.

\clearpage

\noindent {\bf Fig. S3.}
Measured spectra to extract the populations of the system density matrix before or after applying the GRAPE shown in (Fig. S2).
(A) $^{13}$C spectrum of the PPS $\left\vert 000 \right\rangle$ as a result of a $[\pi/2]_y$ pulse applied to the $^{13}$C qubit.
The area (integral) of the absorption peak can be regarded as one benchmark in NMR realizations of
quantum computation. (B)-(D) Signals from the $^{19}$F, $^{13}$C and $^1$H qubits after applying the GRAPE pulse and a $[\pi/2]_y$ pulse to each of the three qubits. All the spectra are exhibited on the $^{13}$C channel through SWAP gates. The integration of each spectral peak gives the difference of two particular diagonal elements of the density matrix $\rho'(t_7)$.

\clearpage
\begin{center} Fig. S1: \end{center}
\begin{figure}[h] \centering
\includegraphics[width=11cm]{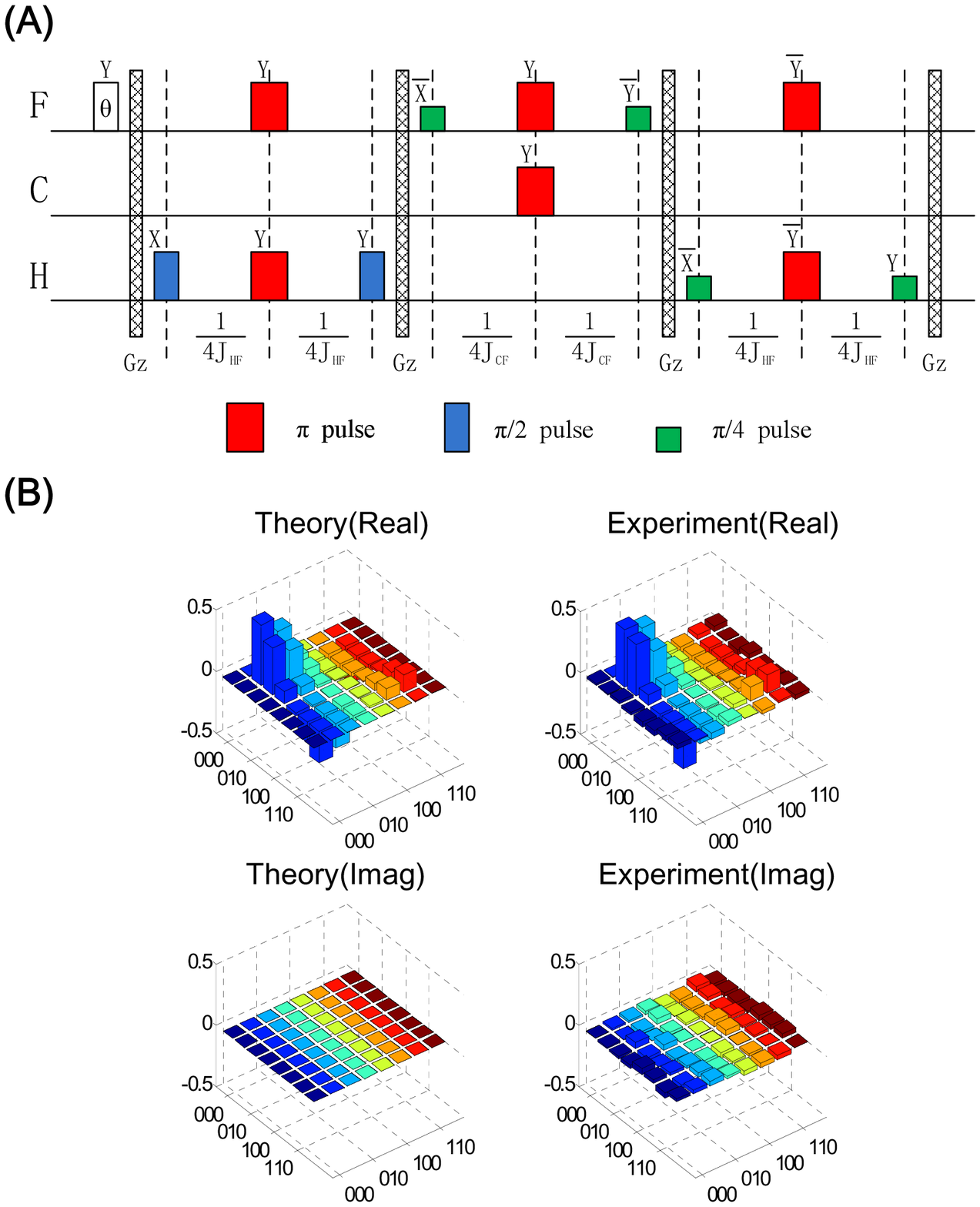}
\label{pps}
\end{figure}

\clearpage
\begin{center} Fig. S2: \end{center}
\begin{figure}[h] \centering
\includegraphics[width=13cm]{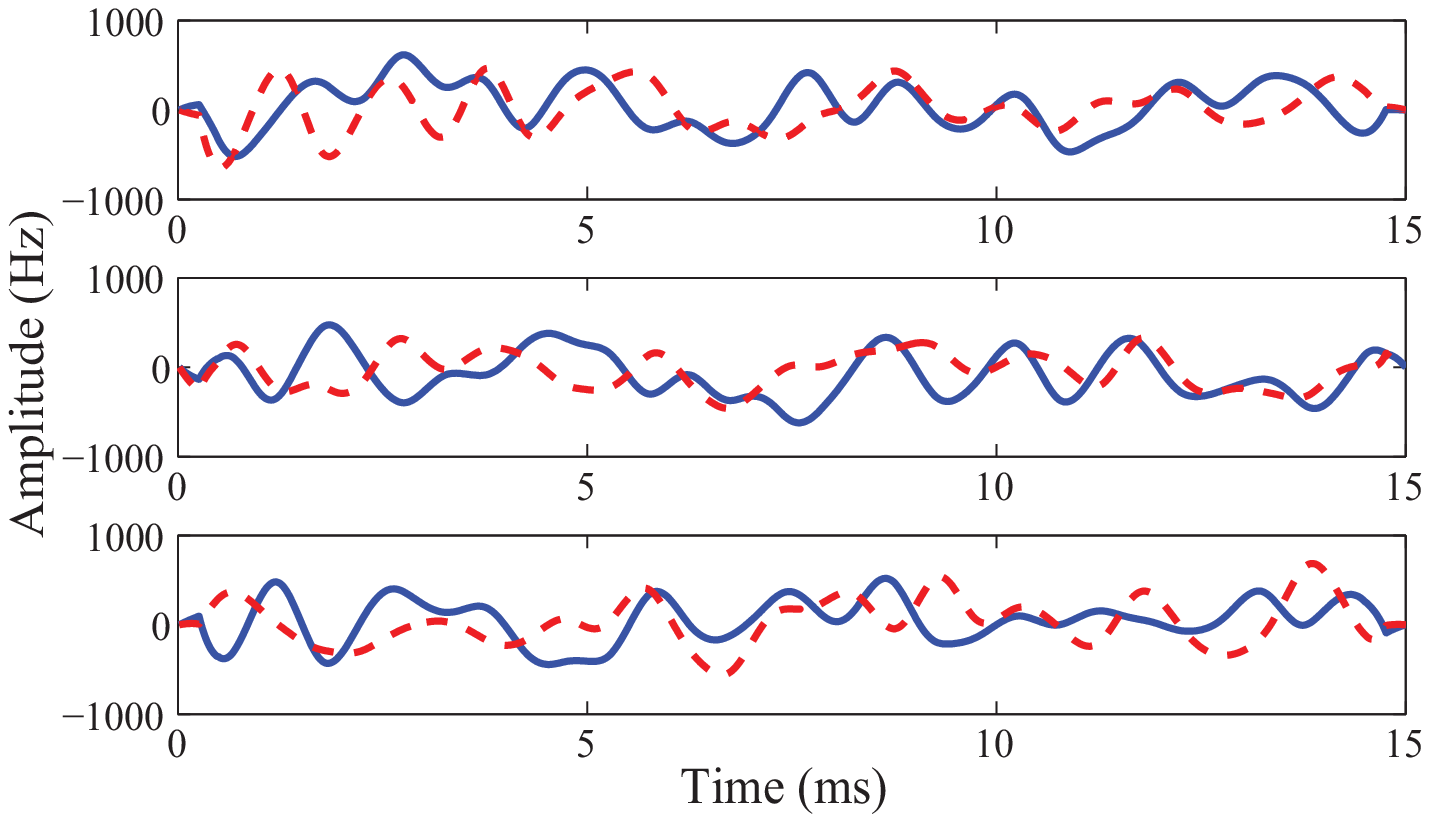}
\label{grape}
\end{figure}

\clearpage
\begin{center} Fig. S3: \end{center}
\begin{figure}[h] \centering
\includegraphics[width=11cm]{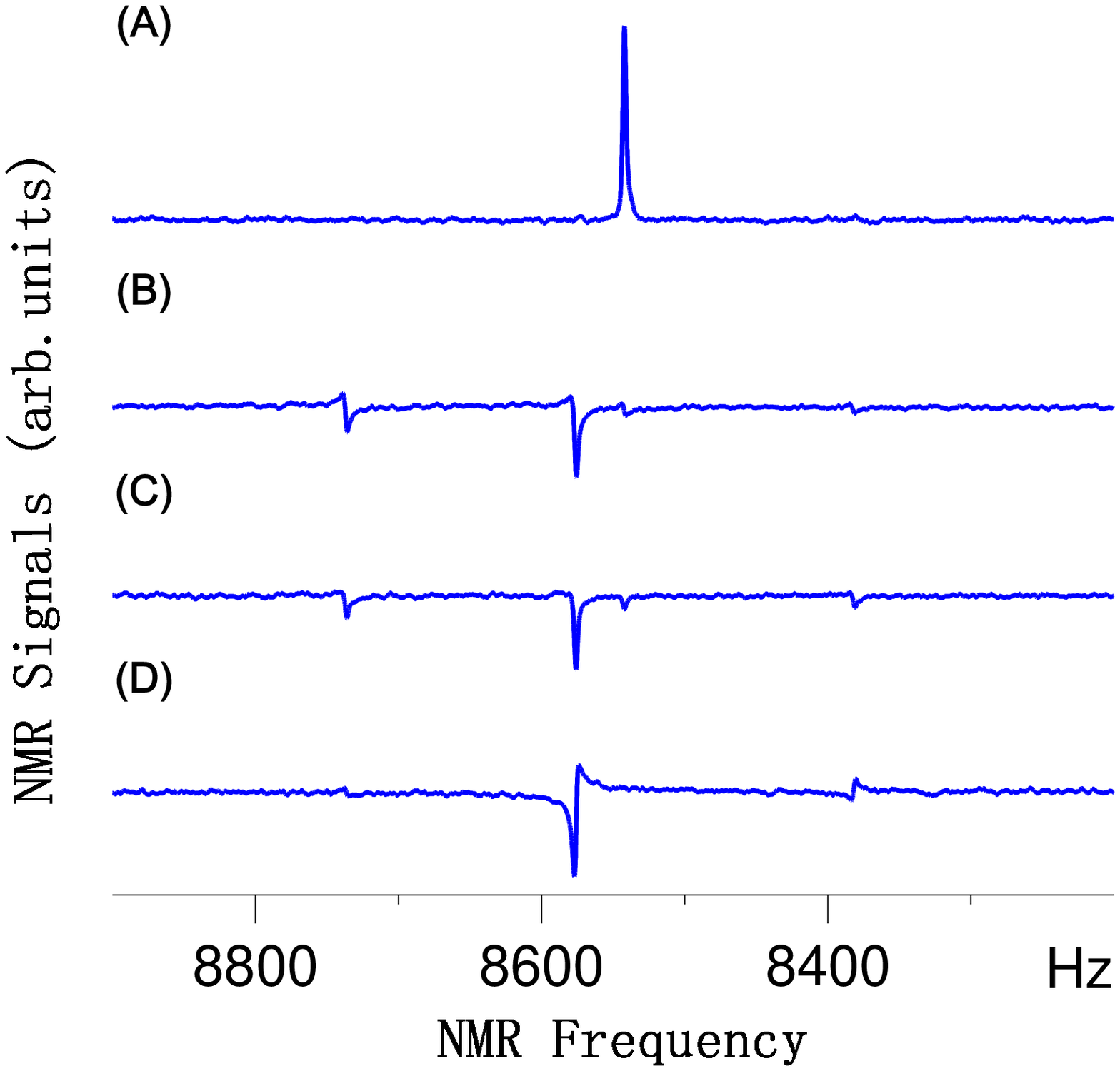}
\label{signal}
\end{figure}

\end{document}